\documentstyle[sprocl,epsf]{article}
\def\be{\begin{equation}}
\def\ee{\end{equation}}
\def\bea{\begin{eqnarray}}
\def\eea{\end{eqnarray}}

\def\gbeta{6/g^2}

\def\xvec{{\vec x}}

\def\Tr{\mathop{\rm Tr}}

\def\Op{\mathop{\cal O}}

\def\LL{\left\langle}	
\def\RR{\right\rangle}	

\def\pbp{\LL\bar\psi\psi\RR}

\def\BE{\begin{equation}}
\def\EE{\end{equation}}
\def\BEA{\begin{eqnarray}}
\def\EEA{\end{eqnarray}}
\def\EL{\nonumber\\}
  
\newcommand{\la}[1]{\label{#1}}

\begin{document}
\rightline{\small AZPH-TH/96-20}
\vglue .5cm

\title{Simulating QCD at finite density with static
quarks\footnote{~To appear in the Proceedings of {\it DPF96} 
University of Minnesota, August 10-15}}
\author{T. Blum}
\address{Brookhaven National Lab, Upton, NY 11973}
\author{J.E. Hetrick and D. Toussaint }
\address{Department of Physics, University of Arizona\\
Tucson, AZ 85721}

\maketitle\abstracts{
We study lattice QCD in the limit that the quark mass and chemical
potential are simultaneously made large, resulting in a controllable
density of quarks which do not move; this is similar in spirit to the
quenched approximation for zero density QCD.  In this approximation we
find that the deconfinement transition seen at zero density becomes a
smooth crossover for any nonzero density at which we simulated, and
that at low enough temperature chiral symmetry remains broken at all
densities.}

Lattice QCD with a nonzero density of quarks is a difficult problem
since the fermion determinant becomes complex\cite{ORIGINAL}, thus we
are led to consider approximations which hopefully capture some of the
essential physics. Here we present a study of QCD at arbitrary quark
density in an approximation where the dynamics of the quarks has been
removed\cite{BHT}. This approach is analogous to the quenched
approximation at zero density.

Our idea is to simultaneously take the limits of infinite quark mass
and infinite chemical potential while the density of quarks remains
fixed. This leaves us with quarks that can be present or absent at
each lattice site, but which do not move in the spatial
directions. The result is a much simpler fermion determinant such that
gauge variables can be easily updated to equilibrium in the background
of a prescribed density of quarks.

With a chemical potential included, the lattice Dirac operator using
Kogut-Susskind quarks is 
\BEA
M(x,y) = 2am_q\delta_{x,y} &+& \sum_{\nu=1,2,3}
                      \Big[ U_\nu(x)\eta_{\nu}(x)\delta_{x+\hat\nu,y}-
U^{\dagger}_\nu(y)\eta_{\nu}(y)\delta_{x-\hat\nu,y}\Big]
\EL
&+& \Big[ e^{\mu a} U_t(x)\eta_{t}(x)\delta_{x+\hat t,y}-
e^{-\mu a} U^{\dagger}_t(y)\eta_{t}(y)\delta_{x-\hat t,y} \Big]\ \
\EEA
Taking limits $m\rightarrow\infty$ and $\mu\rightarrow\infty$
simultaneously leaves $2ma$  along the diagonal, and
the forward hopping terms, $e^{\mu a}U_t$.  Each spatial point
decouples from all others, and the fermion determinant is just a
product of easily computed $SU(3)$ determinants:
\BEA
\la{detMsu3}
\det(M) &=& \prod_\xvec e^{3 \mu a n_t} \det(P_\xvec + C {\bf 1}),\cr
\det(P_\xvec+C) &=& C^3+C^2\Tr P_\xvec + C \Tr P_\xvec^* +1.
\EEA
Here $P_\xvec$ is the Polyakov loop at spatial site $\xvec$, 
and $n_t$ is the number of time slices.
The coefficient of
the unit matrix, $C$, is $(2ma/e^{\mu a})^{n_t}$, and is the
fundamental parameter in our approximation, through which we fix the
density. We are still left with a complex determinant, albeit a much
simpler one, allowing us to generate high statistics.
We estimate expectation values by taking the ratio
\BE\la{op_over_phase}
\langle {\Op} \rangle = \frac
{\langle \Op e^{i\theta} \rangle_{||}}
{\langle e^{i\theta} \rangle_{||}}
\EE
where $\theta$ is the phase of det($M$) and
$\langle\rangle_{||}$ indicates an expectation value in the ensemble
weighted by the modulus of the determinant.

The physical quark density is obtained from
eq.~\ref{detMsu3}
\BE\la{su3den}
\LL n \RR = \frac{1}{a n_t V}\frac{\partial\ln(Z)}{\partial\mu} = 
\frac{1}{V} \LL \sum_\xvec \frac{ C^2 \Tr P_\xvec + 2C\Tr P_\xvec^* + 3 }
{C^3+C^2 \Tr P_\xvec + C \Tr P_\xvec^* + 1} \RR
\EE
where $V$ is the spatial volume. At $C=\infty$ the density is $0$; at
$C=0$ the system is saturated with density 3 per site; $C=1$
represents ``half-filling'' and the density is $3/2$.

We have run simulations on $6^3\times 2$, $8^3\times 2$, $10^3\times
2$, and $6^3\times 4$ lattices as described in Ref. 2.  The lattice
spacing was set by measuring the rho mass.  In figure 1 we summarize
the behavior of the Polyakov loop magnitude ($|P|$) in the
$T$---$\rho$ plane, on $n_t=2$ lattices, with a fitting function. At
zero density, we see the strong first order transition at $T_c$. As
the density increases, this transition becomes a smooth crossover for
all nonzero values of the density at which we simulated (our smallest
density just below the transition region was $\sim$ 0.02
quarks/fm$^3$). We are currently investigating the way in which the
first order transition disappears by examining very low densities near
the transition.
Since we see no systematic dependence of the crossover on
the spatial size, except for the expected decrease of the Polyakov
loop magnitude on cold lattices, we conclude that this rounding is not
a finite size effect.

\begin{figure}[htb]
\epsfxsize= 10cm
\epsfysize= 5cm
\epsffile[0 280 448 570]{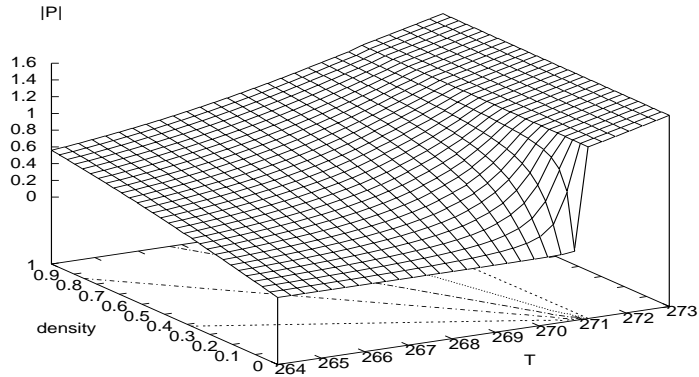}
\caption{Polyakov loop magnitude versus density (q/fm$^3$) and
temperature (MeV). This surface is a smooth function fit to our data.}
\label{fig:1}
\end{figure}

We use $\pbp$ evaluated for light quarks as an indicator of chiral
symmetry breaking in the presence of a finite density of massive
quarks. In a series of runs on $6^3\times 4$ lattices, we began with
$\gbeta=5.0$, which is a fairly cold lattice, at a temperature less
than half the zero density $T_c$.  We find that $\pbp$ remains large
at this (cold) value of $\gbeta$ even up to a density of 1.5
quarks/site (where $C=1$).  Maintaining this maximum density (1.5
q's/site) and increasing the temperature, we find a crossover to
restored chiral symmetry at $\gbeta\approx 5.3$, a significantly lower
temperature than the zero density $n_t=4$ transition, which occurs at
$\gbeta
\approx 5.7$.


Does this static approximation have anything to do with real QCD?
Certainly the nature of the high temperature transition at zero
density depends strongly on the presence of dynamical quarks.
However, it is not {\it a priori} clear to us
that a deconfinement transition or chiral symmetry restoration driven
by high density should depend on the quarks moving, or whether the
mere presence of the quarks would be enough.  In particular, we had
not expected to see the zero density first order transition disappear
for very small quark densities, or the signal of chiral symmetry
restoration to vanish. This suggests that we might want to re-examine
the conventional wisdom that a high density of quarks causes a phase
transition similar to that caused by high temperature.  This work was
supported by DOE grant DE-FG03-95ER-40906.

\end{document}